\documentclass[preprintnumbers,amsmath,superscriptaddress,amssymb,prl,twocolumn]{revtex4-1}

\usepackage{graphicx}
\usepackage[usenames,dvipsnames]{xcolor}
\usepackage{epstopdf}
\usepackage{amsmath,amsthm,amssymb}
\usepackage{latexsym}
\usepackage{bm}
\usepackage{dcolumn}
\usepackage{braket}
\usepackage[normalem]{ulem}
\usepackage{times}
\usepackage{hyperref}

\newcommand{\bu}{\mbox{\boldmath$u$}}

\begin{document}
\bibliographystyle{apsrev4-1}

\title{Entropic elasticity and negative thermal expansion in a simple cubic crystal}

\date{\today}
\date{May 24, 2019}

\author{David Wendt}
\altaffiliation[Present address:]{Island Trees High School, New York 11756, USA}
\affiliation{Condensed Matter Physics and Materials Science Division, Brookhaven National Laboratory, Upton, NY 11973, USA}

\author{Emil Bozin}
\affiliation{Condensed Matter Physics and Materials Science Division, Brookhaven National Laboratory, Upton, NY 11973, USA}

\author{Joerg Neuefeind}
\affiliation{Neutron Scattering Division, Oak Ridge National Laboratory, Oak Ridge, TN 37831, USA}

\author{Katharine Page}
\affiliation{Neutron Scattering Division, Oak Ridge National Laboratory, Oak Ridge, TN 37831, USA}

\author{Wei Ku}
\altaffiliation[Present address:]{Department of Physics and Astronomy, Shanghai Jiao Tong University, Shanghai 200240, China}
\affiliation{Condensed Matter Physics and Materials Science Division, Brookhaven National Laboratory, Upton, NY 11973, USA}

\author{Limin Wang}
\altaffiliation[Present address:]{GE Healthcare, Chicago, IL, USA}
\affiliation{Condensed Matter Physics and Materials Science Division, Brookhaven National Laboratory, Upton, NY 11973, USA}

\author{Brent Fultz}
\affiliation{Department of Applied Physics and Materials Science, California Institute of Technology, Pasadena, California 91125}

\author{Alexei Tkachenko}
\affiliation{CFN, Brookhaven National Laboratory, Upton, New York 11973, USA}

\author{Igor Zaliznyak}
\email{Corresponding author: zaliznyak@bnl.gov}
\affiliation{Condensed Matter Physics and Materials Science Division, Brookhaven National Laboratory, Upton, NY 11973, USA}

\begin{abstract}
{
While most solids expand when heated, some materials show the opposite behavior: negative thermal expansion (NTE) \cite{Barron_book1999}. In polymers and biomolecules, NTE originates from the entropic elasticity of an ideal, freely-jointed chain \cite{Flory_book1969,deGennes_book1979}. The origin of NTE in solids has been widely believed to be different \cite{Sleight_AnnuRev1998,Barrera_etal_JPCM2005,Lind_Materials2012,Dove_RepProgPhys2016}.
Our neutron scattering study of a simple cubic NTE material, ScF$_3$, overturns this consensus.
We observe that the correlation in the positions of the neighboring fluorine atoms rapidly fades on warming, indicating an uncorrelated thermal motion constrained by the rigid Sc-F bonds.
This leads us to a quantitative theory of NTE in terms of entropic elasticity of a floppy network crystal, which is in remarkable agreement with experimental results. We thus reveal the formidable universality of the NTE phenomenon in soft and hard matter. 

}
\end{abstract}
\maketitle

Near zero, or negative thermal expansion is well known in metallic alloys of the invar (Fe$_{0.64}$Ni$_{0.36}$) family, where it is closely related to electronic magnetism \cite{Guillaume_CompRend1897}. These alloys are widely used in applications requiring dimensional stability of metallic parts, e.g. in precision instruments, watches, and engines. Until recently, much less attention was paid to insulating NTE ceramics, which hold promise for numerous applications in electronics, optics and medicine \cite{Sleight_AnnuRev1998,Barrera_etal_JPCM2005,Lind_Materials2012,Dove_RepProgPhys2016}. Somewhat surprisingly, thanks to the specific crystal lattice geometry, NTE in these materials can have the same physical origin as a more common, positive thermal expansion: atomic thermal motion.

Interest in such systems was renewed with the observation of large isotropic NTE in zirconium tungstate, ZrW$_2$O$_8$, and then in the structurally related AM$_2$O$_8$, AM$_2$O$_7$ and A$_2$M$_3$O$_{12}$ phases (A = Zr, Hf, Sc, Y, ... and M = W, V, Mo, P, ...) and their solid solutions \cite{Sleight_AnnuRev1998,Mary_Science1996}, which opened avenues for designing ceramic materials with tailored thermal expansion \cite{Sleight_AnnuRev1998,Barrera_etal_JPCM2005,Lind_Materials2012,Dove_RepProgPhys2016}. These compounds have complex crystal structures, which can be viewed as three-dimensional (3D) networks of AX$_6$ octahedra and MX$_4$ tetrahedra (X = O) that share the corner X atoms and, most importantly, contain nearly straight, two-fold-coordinated M-X-M and M-X-A linkages (the so-called open framework structures \cite{Sleight_AnnuRev1998}).

NTE in such a structure can be explained by the transverse thermal motion of anion atoms, X, in the presence of the strong M-X bond, which has small, or negligible thermal expansion: the so-called ``guitar string effect'' \cite{Sleight_AnnuRev1998,Hu_JACS2016,Ernst_Nature1998}. As the amplitude of the anion transverse vibration increases with temperature, the metal atoms in M-X-M linkages are pulled closer together, thus causing the net contraction of the structure. While this simple picture does not consider the correlated motion of nearby X anions caused by their interactions in the lattice, in what follows we show that it provides an accurate description of NTE in ScF$_3$ \cite{Greve_JACS2010}. 

\begin{figure}
\includegraphics[width=.45\textwidth]{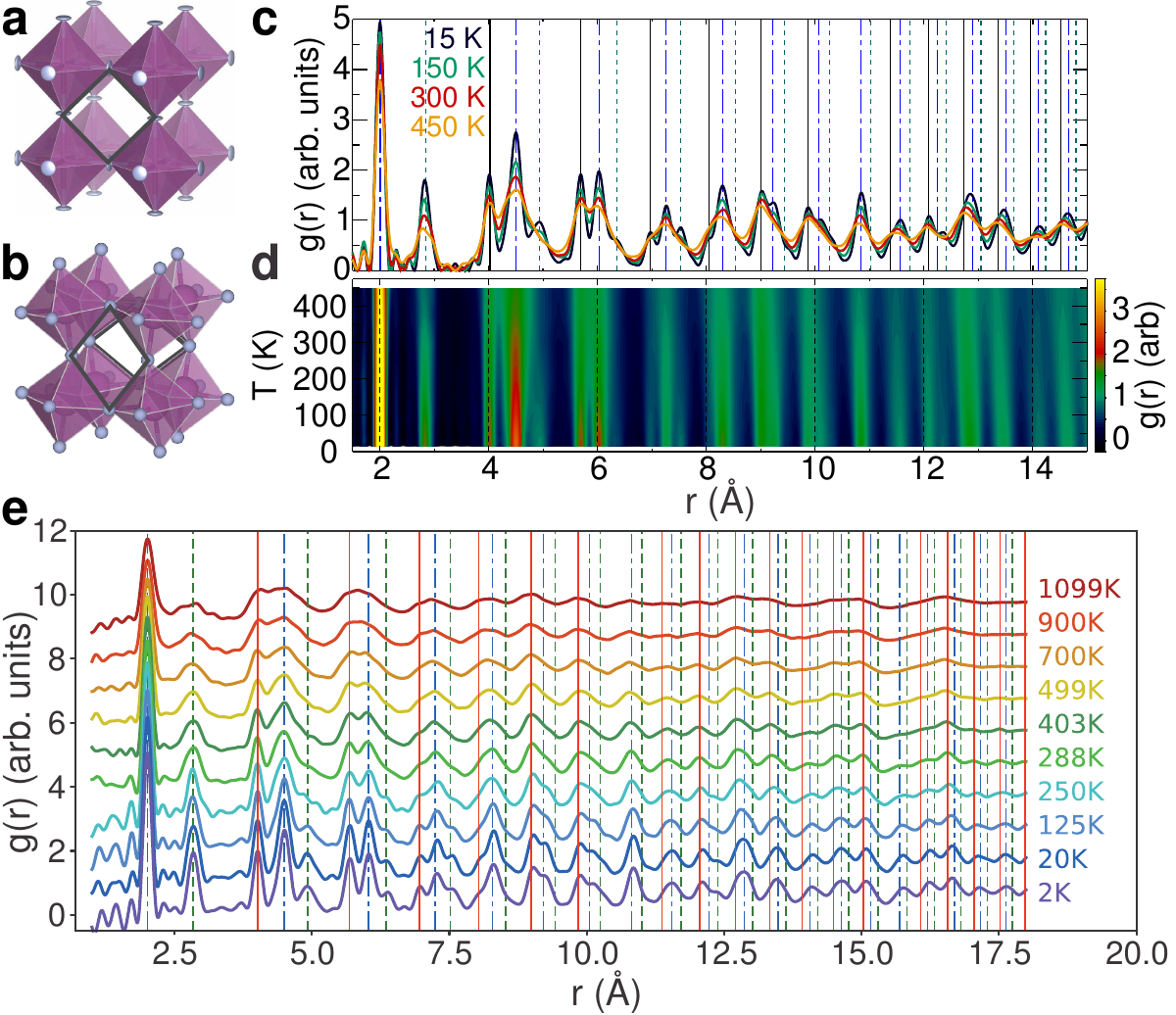}
\caption{
{\bf a} The P$m3m$ cubic perovskite crystal structure of ScF$_3$. Disk-shaped ellipsoids at the vertices of Sc-centered octahedra illustrate large and anisotropic thermal displacements (TD) of fluorine atoms refined at 500 K. {\bf b} The traditional, ball-and-stick representation of the structure, which illustrates the octahedral tilts within the putative rigid unit motion (RUM) model. {\bf c} Pair distribution function, $g(r)$, in 15 K to 450 K temperature range obtained from neutron total scattering measurement on ScF$_3$ powder sample at NPDF using the wave vector range up to $Q_{max} = 27 \AA^{-1}$. {\bf d} The color map representation of the temperature evolution of $g(r)$ emphasizes the negative shift of peaks with increasing temperature, which is most evident at large $r$. {\bf e} PDF measured on the same sample at NOMAD diffractometer for temperatures from 2 K to 1100 K (bottom to top). Here, each curve is an average of $g(r)$ obtained using the wave vector ranges with $Q_{max}$ varying from 23~\AA$^{-1}$ to 32~\AA$^{-1}$ by increments of 1~\AA$^{-1}$. For visibility, data at each temperature above 2 K are shifted upwards by 1. The vertical lines in panels {\bf c} and {\bf e} mark nominal distances corresponding to Sc-F (dash-dotted), F-F (dashed) and Sc-Sc (overlapping with lattice repeats, solid) atom pairs in ScF$_3$ structure. }
\label{Fig1_RDF_struct}
\end{figure}

An appealing model for including anion correlation considers vibrations that preserve the structure of the MX$_n$ polyhedra, which thus move as rigid bodies, without deforming the anion-anion bonds \cite{Sleight_AnnuRev1998,Pryde_JPCM1996,TaoSleight_JSolStChem2003,Tucker_PRL2005,Tucker_JPCM2007}. The relative importance of such rigid unit modes (RUM) has been rationalized by arguing that vibrations distorting the high symmetry of the polyhedron must have a high energy cost and therefore contribute relatively little to NTE \cite{Dove_RepProgPhys2016}. A priori, such an ad-hoc assumption is not required for the NTE effect and its relevance has been a matter of debate \cite{Sleight_AnnuRev1998,Dove_RepProgPhys2016,Pryde_JPCM1996,TaoSleight_JSolStChem2003,Tucker_PRL2005,Tucker_JPCM2007,Cao_PRL2002,Hancock_PRL2004,Schlesinger_PRL2008,Li_PRL2011,Handunkanda_PRB2015,Handunkanda_PRB2016,Young_JMatChem2011}. 
In a structure that is under-constrained, RUM correspond to zero-energy floppy phonon modes \cite{Dove_RepProgPhys2016,Pryde_JPCM1996}. In a fully constrained structure, such as a cubic network of corner-sharing octahedra in ScF$_3$ (Fig.~\ref{Fig1_RDF_struct}a), vibrations that do not distort polyhedra are only present on special low-dimensional manifolds occupying zero volume fraction of the system's phase space
\cite{Dove_RepProgPhys2016,Pryde_JPCM1996}. Nevertheless, it has been argued that phonons in the vicinity of these manifolds, quasi-RUM, which involve only small distortions of the polyhedra have special importance for NTE. In fact, ScF$_3$ was suggested to be a perfect example of an NTE system where the tension effect is enabled by RUM (Fig.~\ref{Fig1_RDF_struct}b) \cite{Dove_RepProgPhys2016}.

Pair distribution function (PDF) analysis \cite{Young_JMatChem2011,EgamiBillinge_book2012} of neutron total scattering is a powerful and direct experimental method for studying average local atomic structures and their relevance for NTE \cite{Dove_RepProgPhys2016,EgamiBillinge_book2012}. The PDF, $g(r)$, which is obtained from the measured scattering intensity, $S(Q)$, gives the probability distribution of inter-atomic distances weighted by the scattering lengths of the constituent atoms and thus is uniquely sensitive to local structural patterns. Figure~\ref{Fig1_RDF_struct}c-e presents the PDF of ScF$_3$ measured on NPDF (c,d) and NOMAD (e) neutron diffractometers at temperatures from 2~K to 1099~K. These measurements are complementary and show good agreement in the temperature range where they overlap.

An inspection of $g(r)$ curves reveals several remarkable features, of which the most important is the distinct behavior of Sc-F and F-F pair distributions. NTE of the average crystal structure is manifested by the systematic negative shift to smaller $r$ of PDF peaks from atomic pairs with large separation, $r$, with increasing temperature. It is most clearly seen in Fig.~\ref{Fig1_RDF_struct}d. The nearest F-F ($\approx 2.8$~\AA) peak shows similar NTE behavior. On the other hand, the nearest-neighbor Sc-F ($\approx 2$~\AA) peak shifts on heating to slightly larger $r$, consistent with the conventional positive thermal expansion (PTE) \cite{Hu_JACS2016,Piskunov_PRB2016}. This peak broadens only moderately with temperature, by about 20\% at 450~K (this accounts for the decrease of peak maximum in Fig.~\ref{Fig1_RDF_struct}), indicating a very stiff Sc-F bond. In contrast, the width of the nearest F-F peak increases markedly, revealing rapid loss of F-F correlation with increasing thermal motion. Even more dramatic is the behavior of further-neighbor F-F distributions. The corresponding peaks (marked by dashed lines in Fig.~\ref{Fig1_RDF_struct}c,e) are only present at T $\lesssim$ 300 K and entirely disappear at higher temperatures, suggesting complete loss of positional correlation between further-neighbor F atoms. Such a liquid-like F-F PDF pattern indicates randomly phased transverse local motion of F atoms and is inconsistent with the RUM model where a large number of F-F distances are constrained by the rigid unit geometry \cite{Dove_RepProgPhys2016,Handunkanda_PRB2015,Handunkanda_PRB2016,Supplementary}.

\begin{figure}
\includegraphics[width=.45\textwidth]{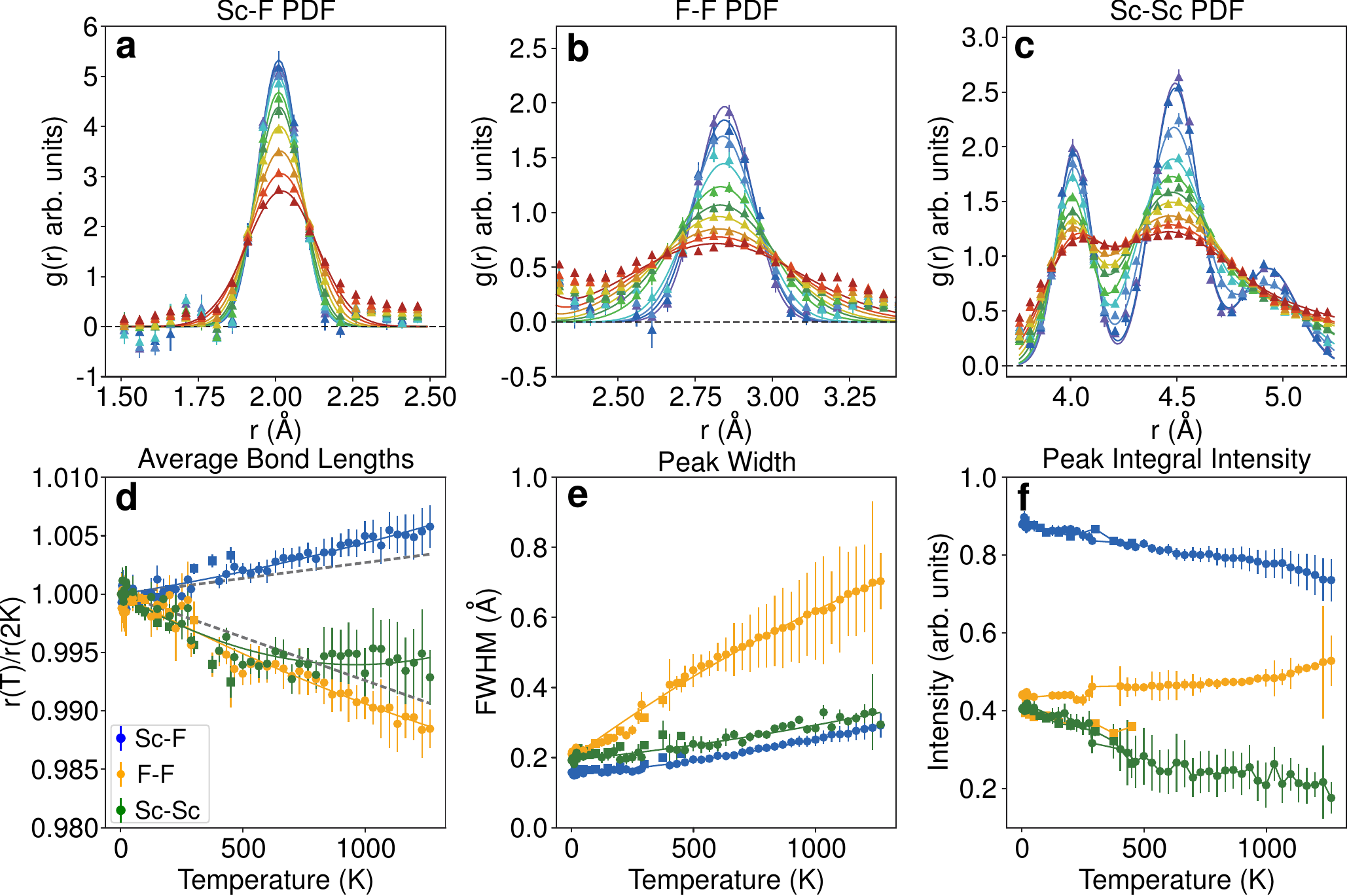}
\caption{
PDF data for Sc-F ({\bf a}), F-F ({\bf b}) and Sc-Sc ({\bf c}, left peak) for the same selected temperatures and with the same color coding as in Fig. 1e (symbols) with Gaussian fits (solid lines). The first two peaks, which correspond to nearest-neighbor Sc-F and F-F pairs can be isolated and were fitted individually with accounting for small intensity overlap, which was either subtracted from the data ({\bf a}), or added to the fit ({\bf b}): a small overlapping intensity contribution of the F-F ($\approx 2.8$\AA) peak to the Sc-F ($\approx 2$~\AA) peak is seen in panel {\bf b}; this contribution has been computed from the model (see text) and subtracted from the data in panel {\bf a}. Due to an overlap of the lattice repeat distance (LRD, $\approx 4$\AA) peak with the next-nearest Sc-F and next-next-nearest F-F peaks, all three were fitted together to a sum of Gaussians, within the data range [3.75\AA, 5.25\AA]. {\bf d} The temperature dependence of the bond lengths obtained from the Gaussian peak position in {\bf a} - {\bf c} shows normal thermal expansion of the Sc-F bond and the contraction of the lattice repeat and the F-F bond. {\bf e} The full width at half maximum (FWHM) of the PDF peaks, reflective of the atomic thermal motion; the broadening of the F-F peak is clearly anomalous. Solid lines in {\bf d}, {\bf e} show fits to quadratic polynominal serving as guides for the eye; dashed lines in {\bf d} are our prediction from entropic elasticity theory for Sc-F and Sc-Sc distances. {\bf f} The integral intensities of the PDF peaks count the participating atoms and nominally should be T-independent; the anomalous loss of the lattice repeat (Sc-Sc) peak intensity indicates loss of the coherent fluorine contribution. Circles are obtained from the NOMAD and squares from the NPDF data. The error bars show one standard deviation accounting for the systematic error; the truncation error in {\bf a} - {\bf c} was estimated by averaging PDFs obtained from the data truncated at different $Q_{max}$, from 23 to 32 \AA$^{-1}$. }
\label{Fig2_fits}
\end{figure}

We quantify the observed behaviors by fitting the first several PDF peaks, which are well resolved and can be uniquely associated with distance distributions of particular atomic pairs, to Gaussian distributions (Fig.~\ref{Fig2_fits}a-c). The results of this analysis are summarized in Figure~\ref{Fig2_fits}d-f. While the Sc-F bond shows PTE of $\approx 5$~ppm at 1000 K, both Sc-Sc, $r_{Sc-Sc}=a$ (LRD, lattice repeat distance), and the nearest F-F, $r_{F-F}$, distances exhibit NTE about twice larger in magnitude (Fig.~\ref{Fig2_fits}d). The large error bars on $r_{F-F}$ reflect dramatic broadening of the F-F peak with temperature. While the full width at half maximum (FWHM) of Sc-F and Sc-Sc (LRD) peaks increase by less than 50\% at 1000 K, the width of the nearest F-F distribution shows nearly an order-of-magnitude larger change, increasing to nearly 1~\AA\ (Fig.~\ref{Fig2_fits}e). This indicates an uncertainty of the nearest F-F distance that is comparable to the $r_{F-F}$ distance itself, clearly invalidating the RUM model assumption of quasi-rigid ScF$_6$ octahedra.

The loss of F-F pair correlation is further revealed by the temperature dependence of the intensity of the LRD ($\approx 4$\AA) peak (Fig.~\ref{Fig2_fits}f). It contains partial contributions from both nearest-neighbor Sc-Sc and next-nearest-neighbor F-F pairs, in proportion $\sigma_{Sc} : 3\sigma_{F} \approx 1.6$, where $\sigma_{Sc}$ and $\sigma_{F}$ are coherent scattering cross-sections of Sc and F, respectively. A PDF peak presents the probability distribution of inter-atomic distance and therefore its integral intensity must be temperature-independent. This roughly holds for Sc-F peak. A small systematic drift of its intensity, which is likely caused by T-dependent background, is within the error bar of the average value. In contrast, the LRD peak rapidly loses a substantial part of its intensity above $\approx$~300 K, where F-F correlations disappear. The decrease is consistent with the loss of the entire $\approx$~40\% partial contribution of F-F pairs, which above $\approx$~300 K contribute to broad background rather than to the narrow LRD peak described by the fit.

\begin{figure}
\includegraphics[width=.45\textwidth]{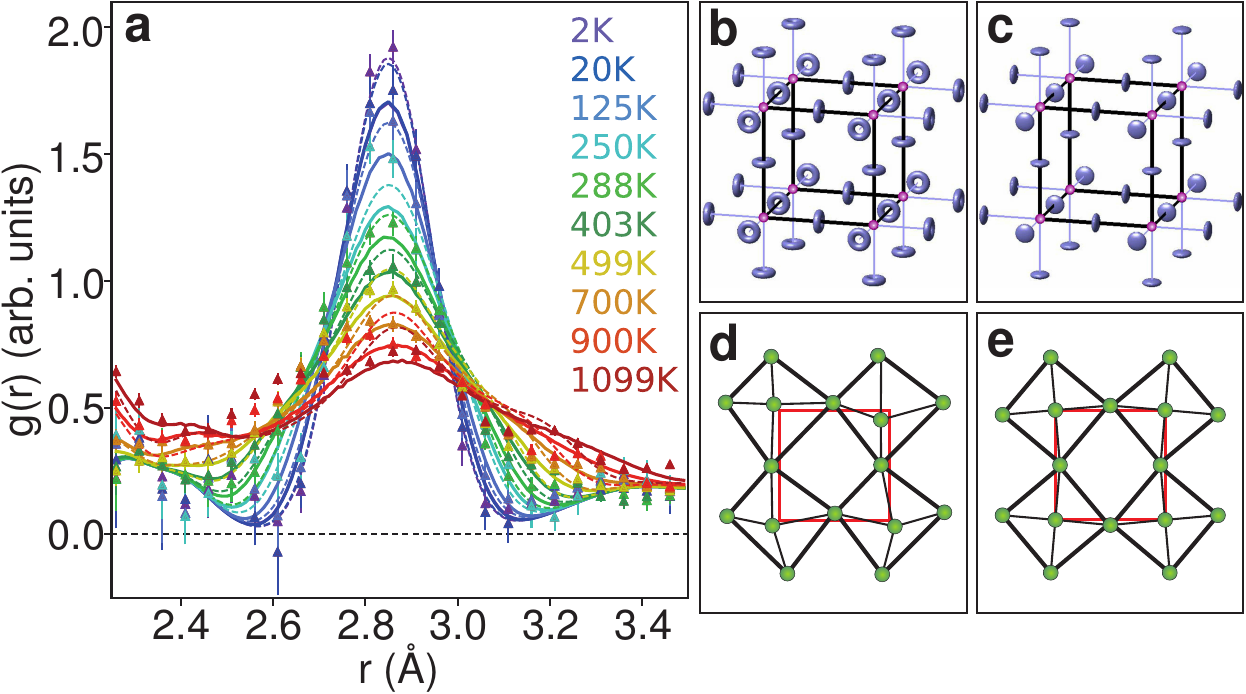}
\caption{
{\bf a} PDF for the nearest neighbor F-F bond (symbols, same as in Fig.~\ref{Fig2_fits}b). The dashed lines show a simplified model for the F-F probability distribution, where positions of the fluorine atoms are randomly distributed on the circles whose radius is determined by the measured thermal displacement parameter obtained from Rietveld refinement. The model was broadened by convolution with the Gaussian of the same width as Sc-F peak at 2~K to account for Sc zero-point motion and truncation effects. The solid lines show an improved model, where circles are replaced by the torus-shaped Gaussian distributions peaked at the same major radius and with the width, represented by the minor radius, which is equal to $1/\sqrt{2}$ of the Sc-F peak width.
The model adequately captures the evolution of the F-F peak position with temperature (Fig.~\ref{Fig2_fits}d).
{\bf b} The ScF$_3$ structure illustrating our model with the tori populated by F thermal motion under the constraint of a rigid Sc-F bond; {\bf c} the traditional representation of the same structure using atomic TD ellipsoids. Both are shown for parameters refined at 500~K. {\bf d} In our model, the entropic motion of F atoms distorts the fluorine octahedra thereby erasing F-F positional correlation, in agreement with the experiment. {\bf e} Opposite to what is observed, rigid octahedra in the RUM model preserve the nearest neighbor F-F bond and also partially preserve the next nearest neighbor F-F correlation, which contributes to the LRD peak.}
\label{Fig3_model}
\end{figure}

\begin{figure}
\includegraphics[width=.45\textwidth]{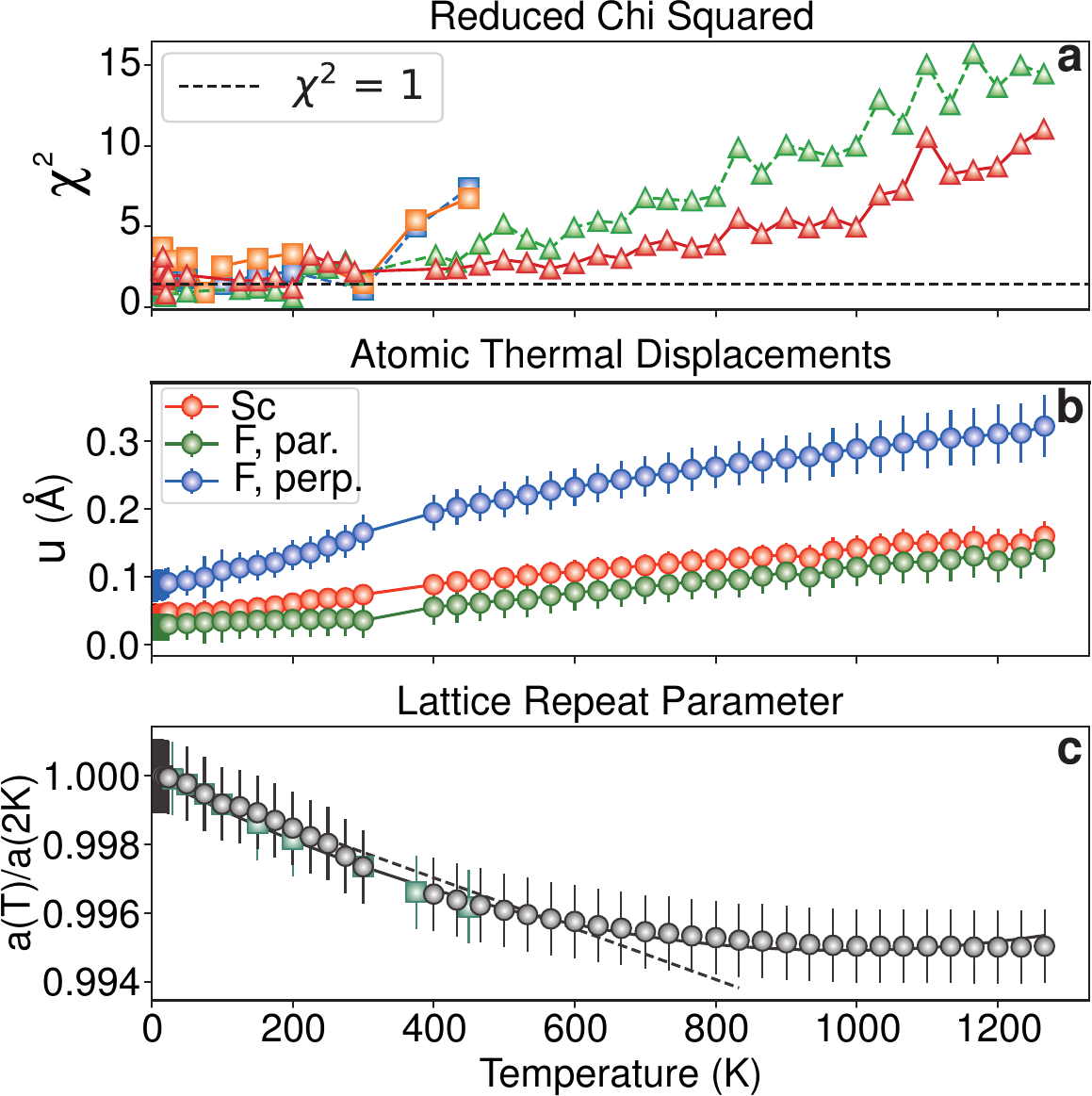}
\caption{
{\bf a} Reduced chi-squared ($\chi^2$) quantifying the accuracy of the entropic model presented in Fig.~\ref{Fig3_model} in describing our NOMAD (triangles) and NPDF (squares) data. Symbols connected by the dashed line show $\chi^2$ for the simplified model with circles in place of tori, which is consistently slightly higher. {\bf b} The atomic TD parameters, and {\bf c} the lattice repeat, $a(T)/a(2{\rm K})$, obtained from the Rietveld refinement of our neutron powder diffraction data, which are used to model the F-F probability distribution. Note that our model has no adjustable parameters: all numbers needed for the simulated F-F probability distribution (the lattice repeat and the F and Sc TD) are obtained from an independent analysis. Where $\chi^2 \sim 1$, the model is indistinguishable from the data. Above $\approx$~600 K, the model begins to fail, with $\chi^2$ reaching $\sim 10$ above 800~K. This is consistent with the failure of its basic assumption of the longitudinally rigid Sc-F bond, which is not unexpected at such high temperatures where the population of the Sc-F longitudinal ($\approx 62$~meV) phonon vibrations becomes significant \cite{Li_PRL2011}. At the same time, the NTE effect fades away ({\bf c}) indicating that its entropic origin is adequately captured by our model. It is also noteworthy that at these high temperatures the Fluorine transverse TD approaches 10\% of the nearest neighbor distance ({\bf b} and Fig.~\ref{Fig2_fits}e), which is close to the Lindemann melting criterion \cite{Barron_book1999}.}
\label{Fig4_Tdeps}
\end{figure}

Motivated by these observations, we use a simple model for the probability distribution of the nearest F-F distance, which is presented in Figure~\ref{Fig3_model}. It assumes un-correlated thermal motion of individual F atoms, which is subject to a single constraint of the rigid Sc-F bond. If Sc atoms were fixed at the nodes of ScF$_3$ lattice, the constraint would result in F atoms following ring trajectories with the Sc-F bond sweeping a cone. The resulting $r_{F-F}$ probability distribution would be that of a distance between two points randomly positioned on the two nearest rings. This model has no adjustable parameters because the radius of the rings, $r_{\perp}$ (the average transverse deviation of F), and $r_{Sc-Sc} = a$ are obtained from the Rietveld refinement of the coherent Bragg scattering contained in our data (Fig.~\ref{Fig4_Tdeps}b,c).
The model can also be set up using Sc-F and Sc-Sc distances refined from PDF peaks (Fig.~\ref{Fig2_fits}), but the accuracy of this refinement is lower.
%
Surprisingly, when broadened by convolution with the Gaussian of the same width as Sc-F peak at 2~K to account for experimental resolution (truncation), our over-simplified model provides adequate description of the measured F-F distribution for all temperatures where NTE is observed (dashed lines in Figs.~\ref{Fig3_model}a and \ref{Fig4_Tdeps}a). In this model, the peak maximum follows lattice NTE, in agreement with Fig.~\ref{Fig2_fits}d. The model can be further improved if instead of rings (or conventional Gaussian TD ellipsoids, Fig.~\ref{Fig3_model}b), F atoms are randomly positioned on a torus-shaped Gaussian distribution peaked at the same major diameter, $2r_{\perp}$, and with the minor diameter representing the F part of the Sc-F peak FWHM (Fig.~\ref{Fig3_model}c). This improved model conjectured by Sleight \cite{Sleight_AnnuRev1998} indeed provides slightly better agreement with the data (Fig.~\ref{Fig3_model}a), as quantified by the reduced mean square deviation, $\chi^2$, presented in Fig.~\ref{Fig4_Tdeps}a. The $\chi^2$ analysis is a standard way to evaluate the goodness of fit: where  $\chi^2 \approx 1$, the data is indistinguishable from the model. With $N \approx 11$ effectively independent data points used in our comparison, $\chi^2 < 3$ places the model within $3\sigma$ interval, or above 99.5\% likelihood level. The model begins to fail above $\approx$~700 K, where $\chi^2$ increases to $\sim 10$, but NTE fades, too. Below $\approx$~200 K the broadening of the F-F peak is small and $\chi^2 \lesssim 1$, which means that within the experimental error our model cannot be distinguished from other models, such as RUM.

The essential implication of our analysis is that thermal motion of even the nearest F atoms is uncorrelated rather than in RUM (Fig.~\ref{Fig3_model}d,e). The spread of the F-F PDF peak with temperature simply follows from the increase in size of the manifold (circle, or torus) populated by each F atom in the course of its thermal motion. Underlying this model is the phenomenon of energy scales separation, where two very different energies govern longitudinal and transverse motion of the F ion.
Then, in some temperature range transverse modes can be thermally excited and the corresponding degrees of freedom be equipartitioned, while longitudinal vibrations are still frozen out. In this case, Sc-F bonds are rigid, while F transverse vibrations are uncorrelated.
Inspection of the vibrational spectra measured in ScF$_3$ \cite{Li_PRL2011,Handunkanda_PRB2015,Piskunov_PRB2016} indeed reveals two major phonon groups, which give rise to maxima in the density of states below $\hbar\omega_t \approx 22$~meV and above $\hbar\omega_l \approx 62$~meV. These correspond to transverse and longitudinal vibrations, respectively. Such separation of energy scales implies that transverse degrees of freedom are thermally excited and equipartitioned at $T  > \hbar\omega_t/k_B \approx 260$~K ($k_B$ is Boltzmann constant), while longitudinal rigidity of Sc-F bond persists up to at least $T \approx \hbar\omega_l/k_B \approx 710$~K. This is exactly the temperature range where NTE is observed and where our model provides very good description of the PDF data (Fig.~\ref{Fig4_Tdeps}).

The exceptional longitudinal rigidity of the Sc-F bond, which underlies the NTE mechanism in ScF$_3$ is rooted in covalence, where the hybridization of Sc and F electronic orbitals that lie deep inside the valence band is responsible for the large energy cost of changing the Sc-F distance \cite{Supplementary,Pauling_book1960,Shaik2016,Teterin_RussChemRev2002,Bocharov_LTP2016}. Such Lewis-type dative bonding where paired electrons delocalize between ions to lower their kinetic energy has recently been described as a ``charge transfer bond'' \cite{Shaik2016}. Although it has long been known that ScF$_3$ is anomalous among supposedly ionic metal trifluorides, MF$_3$ (M = Al, Sc, Fe, In, ...,) \cite{Kury_JACS1959}, only relatively recently has the exceptional strength of the Sc-F bond been traced to the covalent nature of the valence molecular orbitals (MO). The comparative analysis of the X-ray photoemission spectra (XPS) and the density functional theory (DFT) electronic structure calculations \cite{Teterin_RussChemRev2002} for the (ScF$_6$)$^{3-}$ cluster has indicated a large contribution to the Sc-F bonding energy of a specific ($5a_1$) $d-p$ MO at $\approx - 5$~eV. A modest hybridization ($|5a_1\rangle \approx 0.83|F2p\rangle + 0.16|Sc3d\rangle$) does not lead to sizeable charge transfer, such that Sc$^{3+}$ and F$^{-}$ appear to be close to their nominal ionic oxidation states. We performed DFT electronic structure calculations in ScF$_3$, which further support these findings, indicating slightly hybrid valence bands below $\sim - 4$~eV and strong anisotropy of the effective potential of F ions \cite{Supplementary,Bocharov_LTP2016}.

These observations immediately suggest a simple theoretical description of the NTE effect, where each Sc-F bond is treated as a rigid monomer link and the entire ScF$_3$ crystal structure as a floppy network of such freely jointed monomers, a direct 3D analogue of the celebrated model of polymer chains \cite{Flory_book1969,deGennes_book1979} (Fig.~\ref{Fig3_model}b,d). Without electrostatic interactions, the network is under-constrained (floppy): the number of constraints imposed by rigid Sc-F links is 6 per unit cell, while the number of degrees of freedom is 12. In particular, the motion of the Sc ion is constrained by rigid bonds in all 3 directions while each of the F ions has two zero-energy displacement modes corresponding to motion orthogonal to the Sc-F bond. In the absence of external tension the system has no rigidity and would collapse. In ScF$_3$, net Coulomb repulsion between charged ions provides tension (negative pressure), which stabilizes the system and balances its entropic elasticity \cite{Supplementary}.

We thus separate interactions in the system into a sum of the nearest-neighbor pair potentials, which include the cumulative effect of electrostatic Coulomb attraction, core repulsion, and covalent bonding, and in the simplest approximation are treated as rigid links, and the remaining Coulomb potential of non-nearest-neighbor ions. The resulting effective Hamiltonian for the fluorine transverse motion is \cite{Supplementary}
\begin{equation}
\label{H_perp}
H = K + \frac{3N(6-M)e^2}{4\pi \epsilon_0 r} - \sum_i{\frac{\gamma e^2}{4\pi \epsilon_0 r^3}  \frac{\bu_{i \perp}^2}{2}} ,
\end{equation}
where $K$ is kinetic energy, $M \approx 2.98$ is the Madelung constant for ScF$_3$ lattice, $r = a/2$ is half of the lattice repeat, $N$ is the number of sites, $e$ is electron charge, $\epsilon_0$ is vacuum permittivity, $\bu_{i \perp}$ is transverse displacement of the F ion at lattice site $i$, and $\gamma \approx 1.8$ was obtained by lattice summation of Coulomb interactions, similarly to the Madelung constant \cite{Supplementary}. On account of the rigid link constraint, Hamiltonian (\ref{H_perp}) describes the F floppy modes as independent Einstein type oscillators with the frequency, $\hbar\omega_0 = \hbar e \sqrt{\frac{(6-M-\gamma)}{4\pi \epsilon_0 r^3 m_F}} \approx 21.6$~meV. This value compares very favorably with the phonon dispersions measured in ScF$_3$, where low-energy peaks in the density of states are observed below $\hbar\omega_t \approx 22$~meV.

Although the Einstein approximation is expected to perform poorly at low temperatures, where the exact phonon dispersions are important for determining the bulk thermodynamic properties such as heat capacity, or thermal expansion, it works well at $k_B T > \frac{1}{2} \hbar\omega_0$, where the equipartition theorem sets the thermal average of an oscillator Hamiltonian to $k_B T$ per degree of freedom. In this regime, heat capacity obeys the Dulong-Petit law, which is equally well described by both Einstein and Debye models. We thus expect our description, Eq.~(\ref{H_perp}), to be applicable for $T \gtrsim 130$~K. Since each of the F floppy modes has a Hamiltonian of an oscillator with frequency $\omega_0$, the value of $\langle u_\perp^2\rangle = \langle u_{i \perp}^2\rangle$ can be found with the full account for quantum effects and for an arbitrary temperature \cite{Supplementary}. We thus obtain the NTE effect, $\frac{r-r_0}{r_0} \approx - \frac{\langle u_\perp^2\rangle}{2r_0^2} \approx \alpha T$, where $r_0$ is $r$ at $T = 0$ and $\alpha = - \frac{4\pi \epsilon_0 r_0 k_B}{(6-M-\gamma)e^2} \approx -10.1 \cdot 10^{-6}$~K$^{-1}$, in an impressive agreement with the experimental value \cite{Greve_JACS2010} (Fig.~\ref{Fig4_Tdeps}c).

An account for the finite rigidity of the Sc-F bond is done by replacing the rigid link constraint with an interaction potential, $V_b (r_b ) \approx V_b (r_0 ) + f_0 (r_b-r_0 ) + \frac{1}{2} k(r_b-r_0 )^2$. Here, $r_b$ is the bond length, $f_0$ is the tension force that Sc-F linkage provides to compensate the negative electrostatic pressure, and $k$ is the effective harmonic spring constant \cite{Supplementary}. From the measured frequency of the longitudinal Sc-F phonon mode, $\hbar\omega_l \approx 62$~meV, we estimate, $kr_0^2 \approx 26$~eV, in good agreement with our DFT results \cite{Supplementary}. The minimization of the resulting free energy, which includes the entropic term, electrostatics, and bond potential, $V(r_b)$,
with respect to both $r$ and $r_b$, yields the equilibrium values $r=r(T)$, $r_b=r_b (T)$. We thus obtain the relation $\frac{r-r_0}{r_0} -\frac{r_b-r_0}{r_0} \approx \alpha T$, which means that the net entropic tension effect is split between PTE of the Sc-F bond and NTE of the lattice. It further yields the relation, $\frac{r_b-r_0}{r_0} = -\beta \frac{r-r_0}{r_0}$, where $\beta \approx 0.36$, which determines the relative split between the two effects, $\frac{r_b-r_0}{r_0} = \frac{\alpha\beta}{1+\beta}T \approx 2.7 \cdot 10^{-6} T$ and $\frac{r-r_0}{r_0} = \frac{\alpha}{1+\beta}T \approx -7.4 \cdot 10^{-6} T$ \cite{Supplementary}. These predictions are shown by dashed lines in Figs.~\ref{Fig2_fits}d and \ref{Fig4_Tdeps}c, which demonstrate remarkable agreement of our simple theory with experiment. We note that in our estimates we neglected the covalent reduction of the ionic charge on F and Sc ions, which would increase the predicted NTE effect by $\approx 20$\%. This provides a ballpark estimate for the accuracy of our predictions.

We conclude that floppy vibration modes associated with the transverse fluorine displacement in an under-constrained network crystal structure of ScF$_3$ give rise to both negative thermal expansion of the lattice and positive expansion of Sc-F bond. The latter effect is distinct from the conventional positive thermal expansion based on cubic anharmonism of Sc-F bond potential, $V_b (r_b )$ \cite{Barron_book1999}. Instead, it originates from entropic elasticity via floppy modes and is already present for the harmonic Sc-F bond.
RUM, which are of crucial importance for understanding the stability of cubic crystal structure, appear not to be of primary importance for NTE.

In ZrW$_2$O$_8$, the RUM model was challenged by X-ray absorption fine structure (XAFS) studies \cite{Cao_PRL2002} but later was argued to be consistent with neutron PDF measurements \cite{Dove_RepProgPhys2016,Tucker_PRL2005,Tucker_JPCM2007}. However, for a complex material with more than 3 different atom types PDF analysis has a degree of uncertainty because the measured PDF is a sum of PDFs from all atomic pairs where some features may overlap, meaning that it is difficult to identify individual peaks with specific atomic pairs \cite{Dove_RepProgPhys2016}. This problem is absent in ScF$_3$, which has simple cubic structure with only two atom types. In agreement with the earlier molecular dynamics (MD) simulations \cite{Li_PRL2011}, our present results do not support the presence of RUM, indicating that the only rigid unit is the Sc-F bond. This makes the ScF$_3$ structure an under-constrained 3D analog of a freely jointed polymer chain.
%

Based on our experimental observations, we developed a simple theoretical description of the NTE effect in ScF$_3$, which is rooted in entropic elasticity of an underconstrained floppy network, similar in spirit to the celebrated Flory-deGennes theory of polymer elasticity \cite{Flory_book1969,deGennes_book1979}. Our approach presents a paradigm shift, where instead of focusing on peculiar energetics of low-energy lattice vibrations, such as RUM \cite{Barron_book1999,Sleight_AnnuRev1998,Barrera_etal_JPCM2005,Lind_Materials2012,Dove_RepProgPhys2016,Greve_JACS2010,Hu_JACS2016,Pryde_JPCM1996,TaoSleight_JSolStChem2003,SimonVarma_PRL2001,HeCvetkovicVarma_PRB2010,Occhialini_PRB2017}, these vibrations are approximated by Einstein local phonon modes and the focus is on their entropic contribution to free energy. Not only our results provide clear understanding of the entropic elasticity origin of the NTE effect in the practically important class of materials and temperature range, including at and above room temperature, they also provide an accurate, quantitative, textbook description of NTE, thus opening new avenues for predictive modeling of this effect in solids. \\

\emph{Note.} A recent study (https://arxiv.org/abs/1905.09250) reproducing some of our measurements appeared after our results were presented at the 2019 APS March meeting (http://meetings.aps.org/Meeting/MAR19/Session/B33.8). This study also finds substantial flexibility of ScF$_6$ octahedra inconsistent with RUM and supporting our results.

\begin{acknowledgments}
\noindent
Work at Brookhaven National Laboratory was supported by Office of Basic Energy Sciences (BES), Division of Materials Sciences and Engineering, U.S. Department of Energy (DOE), under contract DE-SC0012704. This research at ORNL's Spallation Neutron Source was sponsored by the Scientific User Facilities Division, Office of Basic Energy Sciences, U.S. Department of Energy. \\
\end{acknowledgments}

%

\end{document}